\documentclass{IOS-Book-Article}

\usepackage{mathptmx}
\usepackage{soul}\setuldepth{article}
\usepackage{longtable,booktabs,array}
\usepackage{multirow}
\usepackage{graphicx}
\makeatletter
\def\maxwidth{\ifdim\Gin@nat@width>\linewidth\linewidth\else\Gin@nat@width\fi}
\def\maxheight{\ifdim\Gin@nat@height>\textheight\textheight\else\Gin@nat@height\fi}
\makeatother
\def\hb{\hbox to 11.5 cm{}}

\begin{document}

\pagestyle{headings}
\def\thepage{}

\begin{frontmatter}              

\title{Modelling Business Agreements in the Multimodal Transportation Domain through Ontological Smart Contracts}


\author{\fnms{Mario Scrocca}\thanks{Corresponding Author. E-mail: \href{mailto:mario.scrocca@cefriel.com}{mario.scrocca@cefriel.com}}},
\author{\fnms{Marco Comerio}},
\author{\fnms{Alessio Carenini}}, and
\author{\fnms{Irene Celino}}

\address{Cefriel -- Politecnico di Milano \\ Viale Sarca 226, 20126 Milano, Italy \\
\email{name.surname@cefriel.com}}

\begin{abstract}
The blockchain technology provides integrity and reliability of the information, thus offering a suitable solution to guarantee trustability in a multi-stakeholder scenario that involves actors defining business agreements. The Ride2Rail project investigated the use of the blockchain to record as smart contracts the agreements between different stakeholders defined in a multimodal transportation domain. Modelling an ontology to represent the smart contracts enables the possibility of having a machine-readable and interoperable representation of the agreements. On one hand, the underlying blockchain ensures trust in the execution of the contracts, on the other hand, their ontological representation facilitates the retrieval of information within the ecosystem.
The paper describes the development of the Ride2Rail Ontology for Agreements to showcase how the concept of an ontological smart contract, defined in the OASIS ontology, can be applied to a specific domain. The usage of the designed ontology is discussed by describing the modelling as ontological smart contracts of business agreements defined in a ride-sharing scenario.
\end{abstract}

\end{frontmatter}
\markboth{Scrocca et al., Modelling Ontological Smart Contracts (2022)\hb}{Scrocca et al., Modelling Ontological Smart Contracts (2022)\hb}

\section{Introduction}
In the context of multimodal transportation, a wide set of stakeholders should cooperate to provide passengers with a seamless travel experience. The Shift2Rail Innovation Programme 4\footnote{Shift2Rail IP4, \url{https://rail-research.europa.eu/research-development/ip4/}} (IP4) investigated the design and implementation of an ecosystem of transportation stakeholders relying on an \emph{Interoperability Framework} to support the communication among them and the definition of new services for the users. In this context, the Ride2Rail project\footnote{Ride2Rail, \url{https://ride2rail.eu/}}, 
focused its attention on the integration of ride-sharing alternatives. 

The introduction of new transportation modes, such as ride-sharing, poses additional challenges related to a business environment blending companies and private actors. For this reason, Ride2Rail investigated the usage of blockchain technology to offer suitable guarantees for trust in a multi-stakeholder scenario involving the definition of business agreements. Ride2Rail developed a specific software module for the definition of business agreements as smart contracts, i.e, executable software applications that implement self-executing logic in a blockchain. A smart contract could be implemented as a tool to automate the execution of an agreement that will run when certain conditions are met.

To foster the semantic interoperability of the developed module within the IP4 ecosystem, we designed and adopted an ontology to model the defined agreements through a shared semantic and decoupling the specification of the agreements from its technological implementation. This paper describes the development of such ontology and its application within the Ride2Rail project to describe business agreements in the considered ride-sharing scenario. The main contributions of the paper are: (i) analysis of the literature surveying approaches for an ontological representation of smart contracts, (ii) design and development of an ontology for the definition of business agreements in a multimodal transportation scenario, and (iii) validation of the ontology and exemplification of its usage.

The developed \emph{Ride2Rail Ontology for Agreements} leverages the concept of \emph{Ontological Smart Contract} defined in the OASIS ontology~\cite{cantone_ontological_2021} and investigates how it can be extended to model the semantic of business agreements. Although we focused our work on the transportation domain, the proposed approach can be generalised to model ontological smart contracts in different domains.

The remainder of the paper is structured as follows: Section~\ref{preliminaries} discusses preliminaries regarding the context, terminology and methodology followed; Section~\ref{related-work} frames the work considering the state-of-the-art; Section~\ref{ontology} describes the ontology engineering process and the implemented ontology; Section~\ref{ride2rail-agreements} exemplifies the usage of the ontology reporting how it is used to describe the agreements defined within the Ride2Rail project; Section~\ref{conclusions} draws the conclusions, discusses how the proposed approach can be generalised and future work.

\section{Preliminaries}\label{preliminaries}

This section presents a preliminary introduction to the context and terminology analysed in the paper and the methodology adopted.

\subsection{Context}\label{context}

The Ride2Rail project investigated the specific requirements of ride-sharing for its integration in the multimodal transport ecosystem defined in IP4. The idea behind Ride2Rail is to consider each driver (i.e., a user offering a shared ride with her/his car) as a private transport service provider (TSP) offering transportation services on a specific route. The challenges of such an integration are multiple, for example, the need for a user application that could allow passengers to become a driver offering a ride with their own vehicle, or the need for a dynamic update of shared rides published by drivers for the multimodal journey planning. Moreover, a major challenge is related to how to guarantee trust in an environment where private actors can offer, along with companies, a paid transportation service to passengers.
For this reason, the \emph{Agreement Ledger Module} was designed and developed in Ride2Rail exploiting the blockchain technology to guarantee trust in the definition of agreements between parties (TSPs, drivers and passengers).
The \emph{Agreement Ledger Module} is a software module exposing through an API a set of functionalities relying on smart contracts deployed on a distributed ledger for the digital representation and execution of business agreements. The overall design and implementation of the \emph{Agreement Ledger Module} is documented in the project deliverable~\cite{ride2rail_project_d35_2022}.

The objective of the ontology engineering activity described in this paper is not to transpose ontologically the content of the ledger, but to identify and provide a \emph{semantic description} of the contracts implemented so that they can facilitate interoperability within IP4 and their comprehension from IP4 stakeholders.

For this purpose, the \emph{Ride2Rail Ontology for Agreements} aims at identifying a set of classes and properties to represent the agreements between parties that could be implemented through the \emph{Agreement Ledger Module}. In this paper, we demonstrate how the following agreements could be described using the ontology:
\begin{itemize}
\item The \emph{Ridesharing Booking}, as an agreement between a driver and a passenger.
\item The \emph{Incentive}, as an agreement between different parties to grant, according to a set of conditions, a reward that could promote more sustainable transportation alternatives.
\end{itemize}

\subsection{Terminology}\label{terminology}

In the IP4 scenario, a user with a mobility need can interrogate an application for passengers that is able to process a mobility request and return a set of multimodal offers to cover the itinerary between the required origin and destination.
Each \emph{offer} is associated with a \emph{trip} and a set of \emph{offer items} that the passenger can book to be entitled to travel according to the proposed trip. In the general case, a trip is composed of multiple \emph{travel episode}s, i.e., multimodal trip legs, offered by different \emph{travel expert} systems, provided by different operators and combined to generate a trip.

In this context, the following terminology is introduced for the considered scenario:
\begin{itemize}
\item a \emph{lyft} is defined as a ride-sharing leg in the
  (multi-modal) trip of a passenger, therefore, it is a travel episode for ridesharing;
\item a \emph{ride} is the transportation service offered by a
  \emph{driver} that enables a travel episode for one or more passengers;
\item a \emph{ridesharing booking} represents the booking made by a passenger of an offer item associated with a ride offered by a driver;
\item a \emph{Crowd-based Travel Service Provider} is a travel expert system handling offer items for the rides offered by a set of drivers;
\item an \emph{incentive} is an agreement between parties that is offered by an \emph{incentive provider} defining the \emph{incentive mechanism} and \emph{incentive conditions} for granting it.
\end{itemize}

A more detailed discussion of the presented terminology can be found in the Ride2Rail deliverables~\cite{ride2rail_project_d21_2020, ride2rail_project_d24_2021, ride2rail_project_d31_2021}.

\subsection{Methodology}\label{methodology}

The methodology adopted for the definition of the ontology is based on \emph{Linked Open Terms} (LOT)~\cite{poveda-villalon_linked_2019}, a consolidated industrial method to develop ontologies and glossaries. The LOT methodology is divided into four steps: ontology requirements specification, ontology implementation, ontology publication and ontology maintenance. In the following paragraphs, we briefly discuss the first three steps of the methodology performed to design and implement the \emph{Ride2Rail Ontology for Agreements}.

The ontology requirements specification consists of the definition of the ontology requirements considering the purpose and scope of the ontology, domain analysis and an investigation of the existing data flows. The activity starts with the identification of a set of \emph{use cases} and \emph{user stories} for the ontology. A \emph{use case} should answer the following questions: \emph{Who will be the actors interested in querying the ontological data? What are the
expected usages of data modelled through the ontology?}

Considering the \emph{use cases} and \emph{user stories} defined and the domain analysis performed, then the ontological requirements are specified in the form of \emph{competency questions} and \emph{facts}. The set of \emph{competency questions} defines, in the form of hypothetical questions associated with a \emph{user story}, the information that should be possible to retrieve from data modelled using the ontology. The set of \emph{facts} describes the semantics and the requirements associated with the domain-specific terminology (e.g., attributes describing a specific term, etc.). In this phase, domain experts and stakeholders, are involved to ensure a comprehensive set of ontological requirements is specified.

The second step is the implementation of the ontology.
Considering the requirements, a first conceptual model is produced with the required set of classes and properties. Then, in line with the best practices of ontology engineering, the relevant and already existing vocabularies are analysed to assess the possibility of reusing them.

Finally, the actual ontology is coded in the OWL language and it is validated with the support of automatic diagnosis tools and by manually assessing it with respect to the ontological requirements specified.

The third step is the documentation and publication of the ontological model. 

\section{Related work}\label{related-work}

Blockchain-based solutions and Semantic Web technologies are complementary and may benefit each other. Many researchers are investigating how to combine these technologies and also the European Commission is supporting this research area. As an example, we cite the H2020 ONTOCHAIN\footnote{\url{https://ontochain.ngi.eu/}} project financed under the Next Generation Internet initiative.

Different approaches combining these two areas are reported in the literature. J. Cano-Benito et al. discuss in~\cite{cano-benito_towards_2019} six different scenarios: (a) blockchain with semantic meta-data, (b) blockchain with RDF content, (c) blockchain and virtual RDF service to publish its content, (d) blockchain with external pointers to RDF data, (e) blockchain referencing another blockchain through RDF, (f) semantic blockchain implemented relying on Semantic Web.

The implementation of the mentioned scenarios relies on the definition of ontologies to model the content of the blockchain. In Ride2Rail, we were interested in investigating how smart contracts implemented on the blockchain could be described using an ontology. This approach addresses two interoperability needs: (i) the description in an implementation-independent way of the smart contracts defined according to a specific blockchain-based solution, and (ii) the adoption of proper and shared terminology to describe domain entities and their relationships. Moreover, even if out of scope for the Ride2Rail project, it can foster the implementation of virtual RDF services to query the content of the blockchain using the ontology (scenario \emph{c.}).

In the following paragraphs, we analyse existing vocabularies that were considered to evaluate their re-use in the implementation of the ontology.

\subsection{Smart Contracts and Ontologies}\label{smart-contracts-and-ontologies}

In the analysis of the literature, we found relevant related work regarding the definition of ontologies for smart contracts.

The paper \emph{Ontologies for Commitment-Based Smart Contracts}~\cite{de_kruijff_ontologies_2017} defines a platform-independent conceptualization of smart contracts, however, as mentioned also by the authors in their conclusions, it represents an initial model that should be refined and evaluated before being finally implemented and published as an ontology.

Kruijff and Weigand in \emph{Understanding the Blockchain Using Enterprise Ontology}~\cite{de_kruijff_understanding_2017} adopt an ontology-based approach to formalise the terminology related to the blockchain, including smart contracts, but the proposed modelling considers a higher level of abstraction and doesn't allow for the detailed description of a specific smart contract. Moreover, the authors mention that the presented ontology is still an initial model to be validated and finalised.

Similarly, the \emph{Ontology for Smart Contracts} proposed by McAdams~\cite{mcadams_ontology_nodate} identifies the basic terminology for a conceptualisation of smart contracts that could aid in implementing formal reasoning over their behaviour. The proposed contribution is, however, not implemented as an ontology yet.

Finally, the paper \emph{Ontological Smart Contracts in OASIS}~\cite{cantone_ontological_2021} defines the concept of Ontological Smart Contract extending the OASIS ontology for agents, systems, and integration of services ~\cite{cantone_towards_2019}. The proposed ontology defines the concept of a smart contract as an entity to define agreements between agents and specify their terms, independently from the specific blockchain implementation. Differently from the other work in the literature, the OASIS ontology is fully implemented in OWL, published online and its usage is documented and exemplified by the authors. Moreover, we contacted the authors that provided us with additional documentation to re-use the vocabulary and confirmed that a plan for the maintenance of the ontology is in place. For these reasons, we decided to adopt this vocabulary as a basis for our work.

In this paper, we discuss how the OASIS ontology can support the modelling of ontological smart contracts considering different blockchain technologies, and how it can be extended and leveraged to model business agreements in different domains. Indeed, the usage of the OASIS ontology is exemplified in ~\cite{cantone_ontological_2021} considering a trading agent selling stocks and the Ethereum platform, while, in Section~\ref{ride2rail-agreements}, we consider agreements in the transportation domain implemented through the Hyperledger Fabric API\footnote{\url{https://github.com/hyperledger/fabric-contract-api-go}}.

\subsection{IP4 Ontologies}\label{ip4-ontologies}

To support the definition of domain-specific classes and properties, and
to support interoperability within IP4, we analysed the current status
of the ontologies for the multimodal transportation domain defined in the context of IP4.

The IP4 ontology is currently undergoing an in-depth process
of modularization and extension~\cite{connective_project_d12_2020} considering already standardized formats (e.g., Transmodel\footnote{\url{https://www.transmodel-cen.eu/}}, OSDM\footnote{\url{https://unioninternationalcheminsdefer.github.io/OSDM/}}, TRIAS\footnote{\url{https://github.com/VDVde/TRIAS}}, GTFS-RT\footnote{\url{https://developers.google.com/transit/gtfs-realtime}} etc.). A preliminary release of the new modules of the IP4 ontology, currently under finalisation is available on Github\footnote{Working repository  \url{https://github.com/oeg-upm/mobility}}. The two already available modules of the IP4 ontologies address the \emph{Transmodel} concepts (Core, Commons, Fares, Facilities and Journeys submodules)~\cite{ruckhaus_applying_2021} and the \emph{Open Sales and Distribution Model} (OSDM) specification to model the booking process.

\section{Ontology for Agreements}\label{ontology}

The objective of the \emph{Ride2Rail Ontology for Agreements} is to provide a conceptualization of the basic terms for the description of the business agreements defined in the multimodal transportation context discussed in Section \ref{context}. The ontology takes into account the terminology (Section \ref{terminology}) and the ontologies already defined within IP4 (Section \ref{ip4-ontologies}) to support interoperability through shared semantics.

The following sections describe the application of the presented methodology (Section \ref{methodology}) for the design, implementation and publication of the \emph{Ride2Rail Ontology for Agreements}.

\subsection{Ontological Requirements Specification}\label{ontological-requirements-specification}

This section describes the ontological requirements identified for the \emph{Ride2Rail Ontology for Agreements}. The collection of requirements leveraged the analysis of the overall requirements defined for the Ride2Rail project and the specific ones identified for the implementation of the Agreement Ledger Module. Furthermore, additional stakeholders from the transportation domain were involved to take into account additional considerations from the project consortium and other IP4 actors.

Two use cases were identified to support the definition of the ontology.\\

\textbf{UC1 -- Dispute Resolution about Ridesharing}\\
\emph{Description}: in case of a dispute between a driver and a passenger regarding a booked ride, the responsible authority wants to access trusted data to resolve it.\\
\emph{Stakeholders}: Driver, Passenger, Authority\\
\emph{Workflow}: The responsible authority analyses the details of the booking agreement between the driver and the passenger obtaining trusted information that can help in solving the dispute.\\

\textbf{UC2 -- Incentives to promote Ridesharing}\\
\emph{Description}: travellers (both drivers and passengers) are given incentives to involve ride-sharing in their multimodal rides.\\
\emph{Stakeholders}: Passenger, Driver, Travel Service Provider\\
\emph{Workflow}: Incentives are represented as ontological smart contracts and can be queried to get information about the conditions and mechanisms of available incentives within IP4.\\

For each use case, different user stories considering the stakeholders involved were identified. Finally, a set of facts and competency questions was defined considering: (i) the use cases and user stories defined, (ii) the analysis of the business agreements modelled on the blockchain, and (iii) the relevant terminology in the considered domain.

Table \ref{tab:cqs} contains the competency questions identified for the two use cases, terms starting with a capital letter are concepts described in facts. The complete list of user stories, facts and competency questions is reported in the ontology repository\footnote{\url{https://github.com/Ride2Rail/agreement-ledger-ontology/tree/main/requirements}}.

\begin{small}
\begin{longtable}[b]{p{0.45\linewidth}|p{0.3\linewidth}|c}
\textbf{Competency Question} & \textbf{Expected Result} & \textbf{Use case} \\
\midrule
\endhead
What is the origin/destination of the Ridesharing Leg offered by
the Driver and booked by the Passenger? & Origin/Destination of the
Ridesharing Leg involved in the Ridesharing Booking & UC1 \\ \hline
What is the price agreed upon between the Driver and the Passenger? & Price agreed for the Ridesharing Leg involved in the Ridesharing Booking & UC1 \\ \hline
What is the number of seats declared by a Driver offering a Ride?
& Number of seats associated with a Ride & UC1 \\ \hline
What are the incentive agreements involving a TSP as an
incentive provider? & Incentives involving a TSP in the agreement &
UC2 \\ \hline
What are the conditions defined for a given Incentive? & Conditions defined for the applicability of the Incentive & UC2 \\ \hline
Is there a tangible good or benefit associated with a given Incentive? & Benefit associated with the Incentive & UC2 \\
\bottomrule
\caption{Competency Questions identified for the Ontological Requirements Specification}
\label{tab:cqs}\\
\end{longtable}
\vspace{-30pt}
\end{small}

\subsection{Ontology Implementation}\label{ontology-implementation}

To support the ontology implementation phase, we adopted the \emph{Chowlk}\footnote{\url{https://chowlk.linkeddata.es/}} notation and converter \cite{feria_converting_2021} that allows building the conceptual model graphically and then to directly obtain a first serialization of the ontology in OWL.

The design of the conceptual model, starting from a glossary of terms extracted from the ontological requirements, went through several iterations considering also the outcomes of the review of already available ontological and non-ontological data formats.

To facilitate the description of the final conceptual model, we first discuss the reused vocabularies to model the ontological smart contracts and the domain terminology. Then, we present the final version of the ontology through the \emph{Chowlk} notation. Using the diagram, we motivate our design decisions by describing the introduced classes and properties and the alignment with the re-used vocabularies.

\subsubsection{Ontological Smart Contracts}\label{ontological-smart-contracts-in-oasis}

The Ontology for Agents, Systems, and Integration of Services (OASIS) is published online\footnote{OASIS ontology
  \url{https://www.dmi.unict.it/santamaria/projects/oasis/oasis.php}} with the namespace
\texttt{http://www.dmi.unict.it/oasis.owl\#}
(\texttt{oasis:} prefix).

An ontological smart contract in OASIS (\emph{oasis:SmartContract}) is modelled defining the set of entries involved in the agreement (\emph{oasis:SmartContractEntry}) and the set of conditionals (\emph{oasis:ConditionalSet}) specifying the terms of the agreement. Agreement instances are modelled through the class \emph{oasis:SmartContractInstance} and they are associated with a specific \emph{oasis:SmartContract}.

A \emph{oasis:SmartContractEntry} can be of class \emph{oasis:SmartContractEntryParticipant}, describing a participant involved in the agreement, or class \emph{oasis:SmartContractEntry\\Value}, describing values involved. Each \emph{oasis:SmartContractEntry} can be described using the property \emph{oasis:refersExactlyTo}, if it refers to a specific individual for each instance of the described agreement, or using the property \emph{oasis:refersAsNewTo}, if it describes an individual through an \emph{oasis:EntryTemplate}. An \emph{oasis:EntryTemplate} allows to specify the features that an \emph{oasis:SmartContractEntry} should have in an \emph{oasis:SmartContractInstance} of the modelled \emph{oasis:SmartContract}.

The terms of the agreements are modelled using \emph{oasis:Conditional}, which represent an implication between an antecedent (\emph{oasis:ConditionalBody}) and consequent
(\emph{oasis:ConditionalHead}). Whenever the conditions specified in the antecedent hold, then the conditions specified in the consequent must also hold. Both an \emph{oasis:ConditionalBody} and an \emph{oasis:ConditionalHead} can specify multiple conditions modelling different \emph{oasis:ConditionalAtom.} All the \emph{oasis:ConditionalAtom} should be satisfied to satisfy the antecedent/consequent. A \emph{oasis:ConditionalAtom} can be described through:
\begin{itemize}
\item
  \emph{oasis:ConditionalSubject}: representing the subject;
\item
  \emph{oasis:ConditionalObject}: representing the object;
\item
  \emph{oasis:ConditionalOperator:} representing actions
  (\emph{oasis:Action}) from subject(s) to object(s);
\item
  \emph{oasis:ConditionalParameter}: representing a parameter of the action described by the operator (the two subclasses \emph{oasis:ConditionalInputParameter} and \emph{oasis:ConditionalOutputParameter} representing an input and an output parameter, respectively);
\item
  \emph{oasis:ConditionalOperatorArgument}: representing operator arguments for a subordinate characteristic of the operator
\end{itemize}

Also in the modelling of conditionals, an \emph{oasis:EntryTemplate} can be leveraged to specify the features that the entities involved in the conditional.

\subsubsection{Domain Terminology}

For the definition ontology, considering the elicited requirements, we focused mainly on the \emph{OSDM} module of the IP4 ontologies. The namespace for the considered ontology is \texttt{https://w3id.org/mobility/\\osdm/core\#}
(\texttt{osdm:} module), but the publication process is not finalised yet. In particular, we reused classes and properties related to the concepts of:

\begin{itemize}
\item
  \emph{osdm:Offer}: defined as ``a response to a customer mobility request as a result of the travel shopping process, it is composed of offer item(s) for service(s) designed to cover each proposed journey, and, optionally, ancillary services''.
\item
  \emph{osdm:Booking:} defined as ``an operational process as
  part of the sales process to commit to a sales transaction binding the
  customer and supplier to the offer''.
\end{itemize}

To complement the set of classes and properties already available in the IP4 ontologies, the MaaSive Glossary and the conceptualisation effort made in Ride2Rail WP2.1 (Deliverable D2.1~\cite{ride2rail_project_d21_2020} and D2.4~\cite{ride2rail_project_d24_2021}) and WP3.1 (Deliverable 3.1~\cite{ride2rail_project_d31_2021}) were taken into account as non-ontological resources to improve the semantic interoperability of entities modelled in our ontology.

\subsubsection{Ride2Rail Ontology for Agreements}\label{ride2rail-ontology-for-agreements}

The implementation of the \emph{Ride2Rail Ontology for Agreements} is based on the decision to reuse the \emph{oasis:SmartContract} as the class to model a business agreement. Our claim is that concept of \emph{oasis:SmarContract} is modelled in the OASIS ontology using a generic approach that can be extended and applied to different domains once identified a suitable vocabulary for the representation of the domain terminology. An \emph{oasis:SmarContract} not only enables the representation of the agreement and the entities involved a \emph{oasis:SmartContractEntry}, but also a detailed model of the terms of the agreements as \emph{oasis:Conditionals}. The main objective of the ontology is to complement the current IP4 ontologies and extend the OASIS ontology providing the needed terminology to model ontological smart contracts in the multimodal transportation domain.

\begin{figure}[h!]
  \centering
  \includegraphics[width=\linewidth]{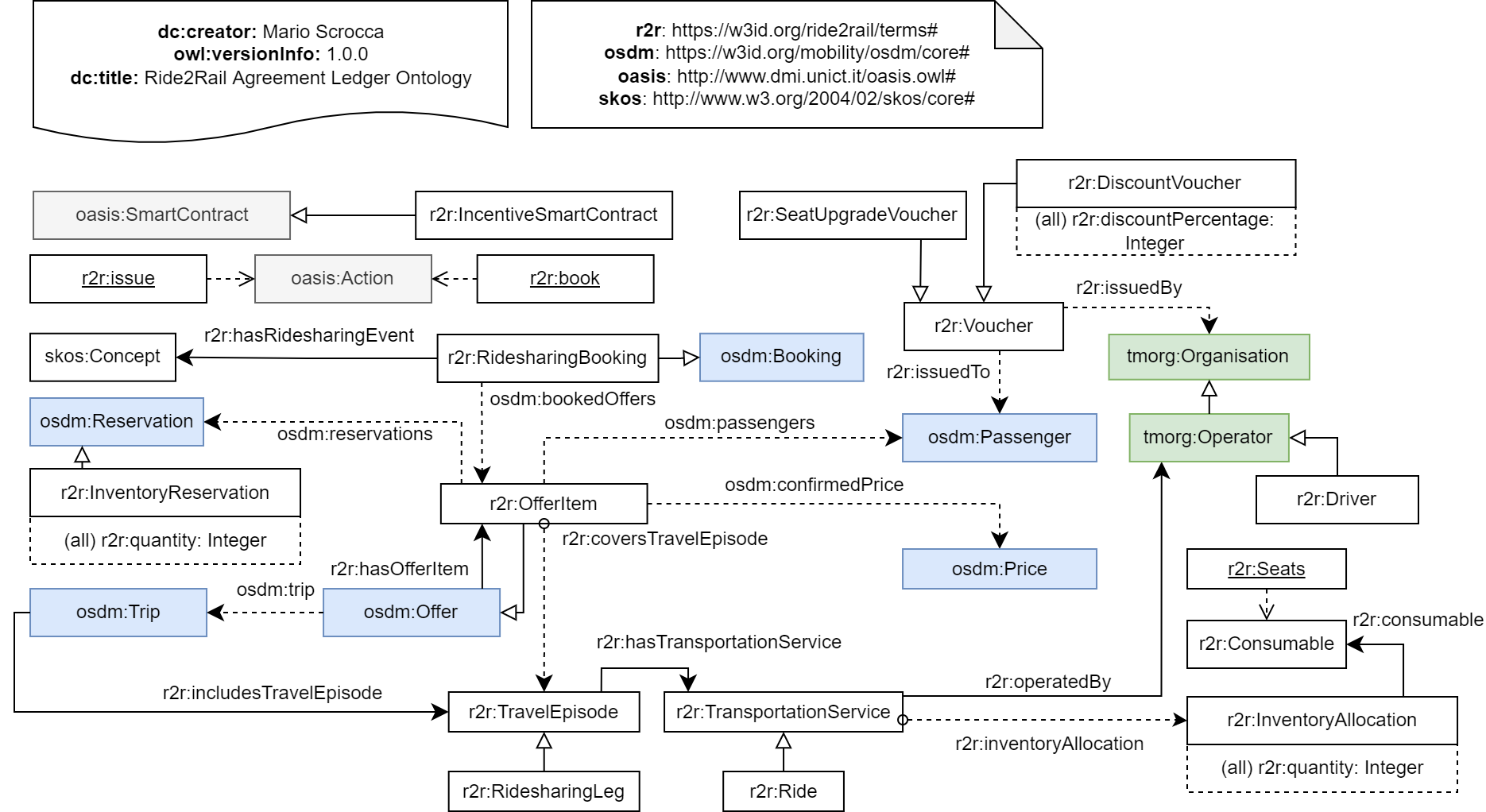}
  \caption{Chowlk diagram for the Ride2Rail Ontology for
Agreements}
    \label{fig:ontology}
\end{figure}

Figure \ref{fig:ontology} adopts the Chowlk notation to describe the classes and properties modelled and their relations with reused vocabularies. The namespace selected for publication is \texttt{https://w3id.org/ride2rail/terms\#} (\texttt{r2r:} prefix).

To support the modelling of the \emph{oasis:SmartContract} for a ridesharing booking, the class \emph{r2r:RidesharingBooking} is defined as a subclass of \emph{osdm:Booking.} An \emph{osdm:Booking} is associated with the booked \emph{osdm:Offer} and the \emph{osdm:Price} paid for the offer. \emph{An osdm:Offer} represents the pair between a computed \emph{osdm:Trip} and the set of products (\emph{osdm:OfferPart}) offered to an \emph{osdm:Passenger} and required to perform the trip.

The ontology extends the concept of \emph{osdm:Offer} defining an \emph{r2r:OfferItem} to model the corresponding term in the IP4 glossary (an offer item is part of an offer, which is provided by a single TSP). As a result, \emph{r2r:RidesharingBooking} is associated with an \emph{r2r:OfferItem} provided by a ridesharing TSP, i.e. the Crowd-based TSP in the ride-sharing scenario.\\
The ontology also defines two classes, \emph{r2r:TravelEpisode} and \emph{r2r:TransportationService} to model, respectively, the IP4 concepts of \emph{travel episode} ("part of a trip operated with the same vehicle") and \emph{transportation service} ("service that provides transportation on a travel episode"). An \emph{osdm:Trip} may include multiple \emph{r2r:TravelEpisode} (possibly aligned with the concept of \emph{osdm:Segment}), an \emph{r2r:TravelEpisode} may be operated by an \emph{r2r:TransportationService}.\\
An \emph{r2r:OfferItem} for an \emph{r2r:RidesharingBooking}, is associated with an \emph{r2r:RidesharingLeg} (subclass of \emph{r2r:TravelEpisode}) that is made possible through an \emph{r2r:Ride} (subclass of \emph{r2r:TransportationService}) operated by an r2r:Driver. In this sense, \emph{r2r:Driver} extends the concept of \emph{tmorg:Operator}.

As specified in the requirements, an \emph{r2r:Ride} may specify an \emph{r2r:InventoryAllocation} that indicates a given quantity of available consumables (\emph{r2r:Consumable}). The ontology defines \emph{r2r:Seat} as an individual of the class \emph{r2r:Consumable}. 
An \emph{r2r:InventoryReservation} (subclass of \emph{osdm:Reservation}) can be associated with an \emph{r2r:OfferItem} defining the number of consumables reserved by the corresponding booking.

To support the modelling of incentives, the class \emph{r2r:IncentiveSmartContract} is defined as subclass of \emph{oasis:SmarContract}. The ontology also models the concept of \emph{r2r:Voucher} to define a redeemable good issued by a \emph{tmorg:Organisation}. Two subclasses are defined for \emph{r2r:Voucher} to model the mechanisms of the incentives implemented in Ride2Rail: a \emph{r2r:DiscountVoucher} allowing to model a certain percentage of discount granted to the beneficiary of the voucher, and a \emph{r2r:SeatUpgradeVoucher} granting an upgrade of seat class for the beneficiary.\\
To express the incentive conditions defined in the requirements, the ontology also defines two \emph{oasis:Action} individuals, \emph{r2r:issue} and \emph{r2r:book}, that can be used as \emph{oasis:ConditionalOperator} in the modelling of conditionals for an \emph{oasis:SmartContract}.

Finally, to model the events that can be associated with a ridesharing booking we decided to implement a SKOS\footnote{\url{https://www.w3.org/2004/02/skos/}} Concept Scheme.

As defined in the requirements, five events are identified in the first level of the taxonomy: \emph{RidesharingStarted} for the start of the ride associated with the ridesharing booking, \emph{RidesharingCompleted} for the completion of the ride associated with the ridesharing booking, \emph{RidesharingCancelled} for the cancellation of the ridesharing booking by the passenger or by the driver, \emph{RidesharingDelayed} for a delay in the ride due to the passenger or to the driver, \emph{RidesharingNoShow} for a passenger or a driver not showing as expected for the booked ridesharing.

The modelled ontology was validated against the ontological requirements and using OOPS! \cite{poveda-villalon_validating_2012} as the state-of-the-art tool for automatic diagnosis of anomalies in the ontology\footnote{A report of the validation is available in the ontology repository \url{https://github.com/Ride2Rail/agreement-ledger-ontology}}.

\subsection{Ontology Publication}\label{ontology-publication}

The \emph{Ride2Rail Ontology for Agreements} is published online following the best practices for ontology publication at
\texttt{https://w3id.org/ride2rail/terms\#} (\texttt{r2r:} prefix). We adopted the \emph{w3id} service for permanent identifiers\footnote{\url{https://w3id.org/}}, and we implemented content negotiation to serve the ontology in different human-readable and machine-readable formats\footnote{Recipe 3 from \url{https://www.w3.org/TR/swbp-vocab-pub}}. The Widoco \cite{damato_widoco_2017} tool was used to generate the ontology documentation, then complemented through diagrams and the description of the main design decisions.
The SKOS taxonomy for ridesharing booking events is published at \texttt{https://w3id.org/ride2rail/rb-events\#} (\texttt{rbe:} prefix) using a similar approach.

The license adopted is the Creative Commons with Attribution right (CC-BY), which allows licensees to copy and distribute the work and make derivative works, giving the authors proper credits.

All the material related to the ontology and the artifacts produced during the ontology engineering process are hosted on Github under the Ride2Rail organisation in the repository \url{https://github.com/Ride2Rail/agreement-ledger-ontology}.

\section{Modelling Business Agreements}\label{ride2rail-agreements}

In this section, we exemplify the usage of the defined \emph{Ride2Rail Ontology for Agreements} discussing the RDF representation of the business agreements implemented through the Agreement Ledger Module in the Ride2Rail project, i.e., the ridesharing booking and the incentives. We decided to distinguish between the ontology and the agreements since the definition of different business agreements (e.g. different conditionals and incentives) is made possible by relying on the implemented ontology.

The RDF dataset is published at \texttt{https://w3id.org/ride2rail/agreements\#} (\texttt{ag:} prefix) and hosted on Github\footnote{\url{https://github.com/Ride2Rail/agreement-ledger-ontology/tree/main/agreements}}. A Chowlk diagram is provided in the ontology documentation for each \emph{oasis:SmartContract}\footnote{The Chowlk diagrams for the defined \emph{r2r:IncentiveSmartContract}s are available also in the repository \url{https://github.com/Ride2Rail/agreement-ledger-ontology/tree/main/docs/diagrams}}. In the following, we discuss the defined agreements and the expected usage of the terms in the considered scenario.

\begin{figure}[h!]
  \centering
  \includegraphics[width=\linewidth]{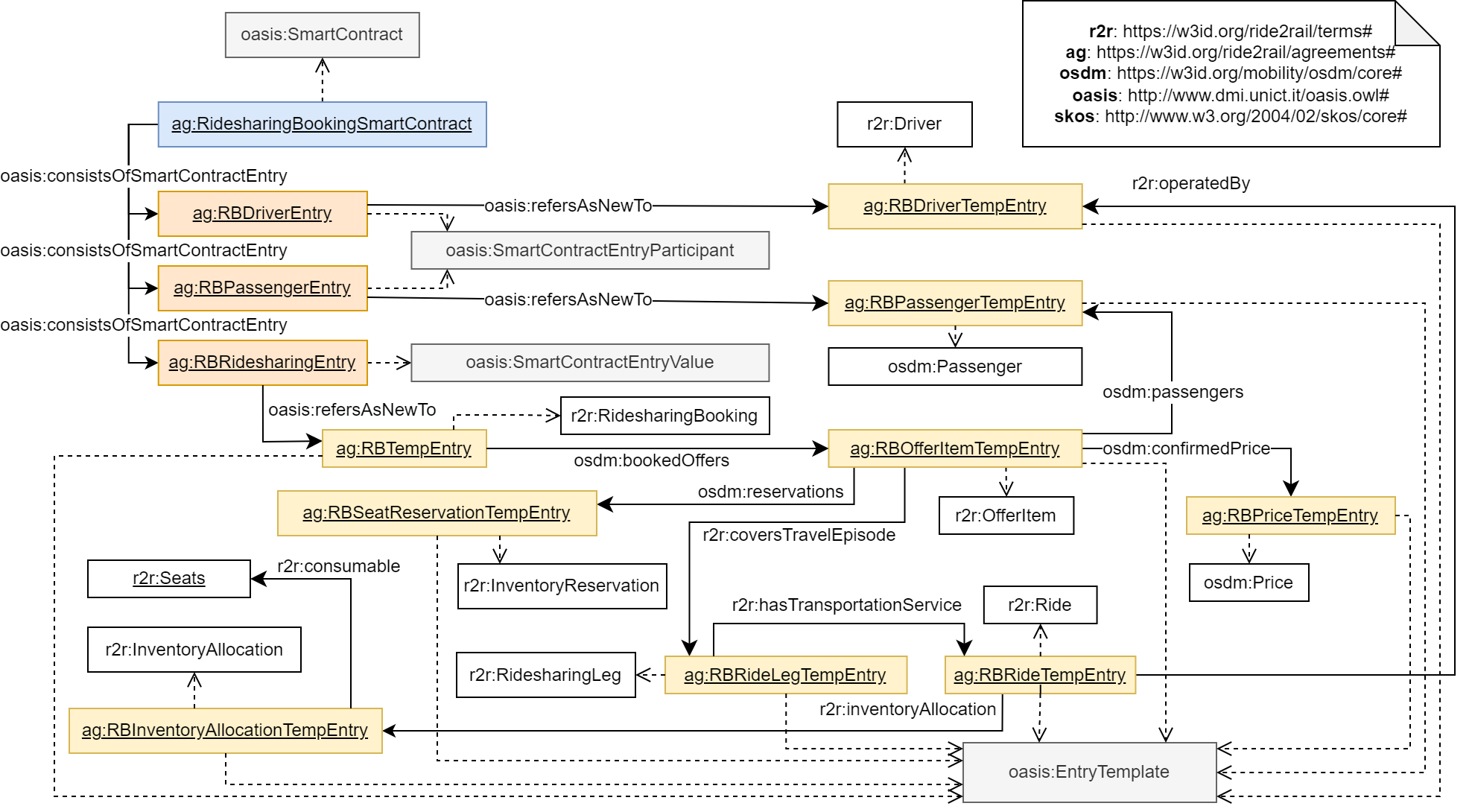}
  \caption{Diagram describing the \emph{RidesharingBookingSmartContract} agreement.}
    \label{fig:rbsm}
\end{figure}

The first agreement is the \emph{ag:RidesharingBookingSmartContract} that defines the entities associated with a ridesharing booking and their relations. Figure \ref{fig:rbsm} represents the Chowlk diagram describing the modelled \emph{oasis:SmartContract}. An \emph{ag:Ridesharing\-Booking\-Smart\-Contract} involves an \emph{r2r:Driver} and an \emph{osdm:Passenger} as participants in an \emph{r2r:RidesharingBooking}. The agreement specifies that the \emph{r2r:Ridesharing\-Booking} is associated with an \emph{r2r:Offer\-Item} for the passenger, that has a specific \emph{osdm:Price}, is associated with an \emph{r2r:RidesharingLeg}, and includes an \emph{r2r:InventoryReservation}. The \emph{r2r:RidesharingLeg} has as transportation service an \emph{r2r:Ride} offered by the driver with a specific \emph{r2r:InventoryAllocation}.

The \emph{ag:RidesharingBookingSmartContract} does not define any \emph{oasis:Conditional} since the terms of the agreement are not directly modelled through the Agreement Ledger Module. Table \ref{tab:conditionals} defines examples of \emph{oasis:Conditional} that can be defined exploiting the ontology, in particular, leveraging the ride-sharing booking events taxonomy. 
Example 1 describes the fact that the money from the passenger should be transferred to the driver if the ride is correctly started and ended, examples 2 and 3 describe potential refund policies if the driver doesn't show up or cancels the ride.

The Agreement Ledger Module implemented in Ride2Rail defines three
agreements related to incentives, i.e., three
\emph{r2r:IncentiveSmartContract}: Ride with other passengers
(\emph{ag:RideWithOtherPassengersIncentive}), Multimodal discount
(\emph{ag: MultimodalDiscountIncentive}), Multimodal repetition discount
(ag:\emph{MultimodalDiscountIncentive3}).\\
Table \ref{tab:incentivesm} summarises the defined agreements reporting the
\emph{oasis:SmartContractEntry} involved and the modelled
\emph{oasis:Conditional}. All the agreements are between an
\emph{osdm:Passenger} and a \emph{tmorg:Operator,} then:
\begin{itemize}
\item
  \emph{ag:RideWithOtherPassengersIncentive} defines a conditional specifying 
  that if the passenger books a ride also booked by another passenger,
  then the TSP issues an \emph{r2r:SeatUpgradeVoucher} to the passenger;
\item
  \emph{ag:MultimodalDiscountIncentive} defines a conditional specifying that if
  the passenger books a multimodal ride involving an
  \emph{r2r:RidesharingLeg} and at least another leg, then the TSP
  issues an \emph{r2r:DiscountVoucher} of 10\% to the passenger;
\item
  \emph{ag:MultimodalDiscountIncentive3} defines a conditional specifying that
  if the passenger books a multimodal ride involving an
  r2r:RidesharingLeg and at least another leg for three times, then the
  TSP issues an r2r:DiscountVoucher of 20\% to the passenger.
\end{itemize}

\begin{footnotesize}
\begin{longtable}[c]{ll|c|c|c}
\multicolumn{2}{l|}{\textbf{Conditional}} &
  \textit{\textbf{example-1}} &
  \textit{\textbf{example-2}} &
  \textit{\textbf{example-3}} \\ \hline
\endfirsthead
\endhead
\multicolumn{1}{l|}{\multirow{2}{*}{\textbf{Entities}}} &
  Participant &
  \begin{tabular}[c]{@{}c@{}}Passenger\\ Driver\end{tabular} &
  \begin{tabular}[c]{@{}c@{}}Passenger\\ Driver\end{tabular} &
  \begin{tabular}[c]{@{}c@{}}Passenger\\ Driver\end{tabular} \\ \cline{2-5} 
\multicolumn{1}{l|}{} &
  Value &
  \begin{tabular}[c]{@{}c@{}}Ridesharing \\ Booking\end{tabular} &
  \begin{tabular}[c]{@{}c@{}}Ridesharing \\ Booking\end{tabular} &
  \begin{tabular}[c]{@{}c@{}}Ridesharing \\ Booking\end{tabular} \\ \hline
\multicolumn{1}{l|}{\multirow{3}{*}{\textbf{Body}}} &
  Subject &
  \begin{tabular}[c]{@{}c@{}}Ridesharing \\ Booking\end{tabular} &
  \begin{tabular}[c]{@{}c@{}}Ridesharing \\ Booking\end{tabular} &
  \begin{tabular}[c]{@{}c@{}}Ridesharing \\ Booking\end{tabular} \\ \cline{2-5} 
\multicolumn{1}{l|}{} &
  Operator &
  \begin{tabular}[c]{@{}c@{}}is associated \\ with\end{tabular} &
  \begin{tabular}[c]{@{}c@{}}is associated \\ with\end{tabular} &
  \begin{tabular}[c]{@{}c@{}}is associated \\ with\end{tabular} \\ \cline{2-5} 
\multicolumn{1}{l|}{} &
  Object &
  Start Event, End Event &
  \begin{tabular}[c]{@{}c@{}}Cancelled \\ Driver\end{tabular} &
  \begin{tabular}[c]{@{}c@{}}No Show \\ Driver\end{tabular} \\ \hline
\multicolumn{1}{l|}{\multirow{5}{*}{\textbf{Head}}} &
  Subject &
  Passenger &
  - &
  - \\ \cline{2-5} 
\multicolumn{1}{l|}{} &
  \begin{tabular}[c]{@{}l@{}}Input \\ Parameter\end{tabular} &
  Price &
  - &
  - \\ \cline{2-5} 
\multicolumn{1}{l|}{} &
  Operator &
  pay &
  refund &
  refund \\ \cline{2-5} 
\multicolumn{1}{l|}{} &
  \begin{tabular}[c]{@{}l@{}}Output \\ Parameter\end{tabular} &
  - &
  Price &
  Price \\ \cline{2-5} 
\multicolumn{1}{l|}{} &
  Object &
  Driver &
  Passenger &
  Passenger \\
\caption{Example of \emph{oasis:Conditional} for a ride-sharing booking smart contract.
}
\label{tab:conditionals}\\
\end{longtable}
\end{footnotesize}

\begin{footnotesize}
\begin{longtable}[c]{ll|c|c|c}
\multicolumn{2}{l|}{\textbf{Incentive Smart Contract}} &
  \textit{\textbf{\begin{tabular}[c]{@{}c@{}}Ride \\ With Other \\ Passengers \\ Incentive\end{tabular}}} &
  \textit{\textbf{\begin{tabular}[c]{@{}c@{}}Multimodal \\ Discount \\ Incentive\end{tabular}}} &
  \textit{\textbf{\begin{tabular}[c]{@{}c@{}}Multimodal \\ Repetition \\ Discount \\ Incentive\end{tabular}}} \\ \hline
\endfirsthead
\endhead
\multicolumn{1}{l|}{\textbf{Entities}} &
  Participant &
  \begin{tabular}[c]{@{}c@{}}Passenger P1\\ Travel Service \\ Provider TSP\end{tabular} &
  \begin{tabular}[c]{@{}c@{}}Passenger P1\\ Travel Service \\ Provider TSP\end{tabular} &
  \begin{tabular}[c]{@{}c@{}}Passenger P1\\ Travel Service \\ Provider TSP\end{tabular} \\ \hline
\multicolumn{1}{l|}{\multirow{4}{*}{\textbf{Body}}} &
  Subject &
  \begin{tabular}[c]{@{}c@{}}Passenger P1, \\ Passenger P2\end{tabular} &
  Passenger &
  Passenger \\ \cline{2-5} 
\multicolumn{1}{l|}{} &
  Operator &
  \begin{tabular}[c]{@{}c@{}}makes \\ a booking\end{tabular} &
  \begin{tabular}[c]{@{}c@{}}makes \\ a booking\end{tabular} &
  \begin{tabular}[c]{@{}c@{}}makes \\ a booking\end{tabular} \\ \cline{2-5} 
\multicolumn{1}{l|}{} &
  \begin{tabular}[c]{@{}l@{}}Operator \\ Argument\end{tabular} &
  - &
  - &
  3 times \\ \cline{2-5} 
\multicolumn{1}{l|}{} &
  Object &
  Ride R1 &
  \begin{tabular}[c]{@{}c@{}}Offer for a \\ Trip involving a \\ Ridesharing Leg \\ and another Leg\end{tabular} &
  \begin{tabular}[c]{@{}c@{}}Offer for a \\ Trip involving a \\ Ridesharing Leg \\ and another Leg\end{tabular} \\ \hline
\multicolumn{1}{l|}{\multirow{3}{*}{\textbf{Head}}} &
  Subject &
  TSP &
  TSP &
  TSP \\ \cline{2-5} 
\multicolumn{1}{l|}{} &
  Operator &
  issue &
  issue &
  issue \\ \cline{2-5} 
\multicolumn{1}{l|}{} &
  Object &
  \begin{tabular}[c]{@{}c@{}}Seat Upgrade \\ Voucher for \\ Passenger P1\end{tabular} &
  \begin{tabular}[c]{@{}c@{}}10\% Discount \\ Voucher for \\ Passenger\end{tabular} &
  \begin{tabular}[c]{@{}c@{}}20\% Discount \\ Voucher for \\ Passenger\end{tabular} \\
\caption{\emph{IncentiveSmartContract}s implemented in Ride2Rail}
\label{tab:incentivesm}\\
\end{longtable}
\end{footnotesize}

The described \emph{oasis:SmartContract}s not only enable the sharing of the modelled business agreements between the IP4 stakeholders, but also the representation of smart contract instances stored on the blockchain to support the use cases defined.

\section{Conclusions}\label{conclusions}

In an ecosystem comprehending various stakeholders, the implementation of business agreements through a distributed ledger provides several benefits regarding information trust and the automatic execution of the agreed terms modelled as smart contracts. This approach, however, doesn't provide any guarantee about the interoperability of the defined agreements from a technological and semantic perspective. On one hand, the domain terminology shared among the involved stakeholders should be referenced by the modelled entities, on the other hand, other software systems can benefit from a  machine-readable representation of the agreements.

The concept of ontological smart contract, defined in the OASIS ontology, supports the representation of business agreements independently from their implementation and relying on standardised vocabularies.
The defined \emph{Ride2Rail Ontology for Agreements} enables the application of this concept to support the interoperability of business agreements in the multimodal transportation scenario considered by the Ride2Rail project. Two use cases were considered: the representation of the ride-sharing booking as an agreement between the driver and the passengers for dispute resolution, and the definition of incentives as agreements between different stakeholders to promote the usage of multimodal transportation.
In the paper, we validated and exemplified the usage of the ontology by modelling the specific business agreements implemented in the project on the ledger. 

The discussed approach can be generalised to support an ontological representation of smart contracts in different domains. The following steps summarise the discussed activities: (i) investigation of the business agreements to be modelled in the considered scenario (use cases and user stories), (ii) analysis of the domain terminology covered by the business agreements (facts and competency questions), (iii) identification of existing vocabularies covering the relevant domain entities and relationships, and/or implementation of an ontology supporting their representation, (iv) modelling of each business agreement as an ontological smart contract identifying the involved entities and the terms of the agreement, and, optionally, (v) representation of specific entities of the business agreement stored on the ledger using the ontology. In this way, different stakeholders are able to access through uniform terminology a description of the smart contracts and, possibly, their instances.

In future work, we will investigate the materialisation/virtualisation of smart contracts and/or related instances from the blockchain to enable querying according to the defined ontology. In particular, we will explore the configuration of semantic conversion pipelines~\cite{scrocca2020turning} and the exploitation of the obtained knowledge graph.
Moreover, we would like to extend the scope of the defined ontology to enable the representation of heterogeneous agreements in the multimodal transportation domain, for example, considering requirements for the sharing and electric mobility~\cite{scrocca_urban_2021}. Finally, the evolution of the suite of IP4 ontologies will be taken into account to extend and update the defined ontology.

\begin{acknowledgements}
The presented research was supported by the RIDE2RAIL project  (Grant Agreement  881825), co-funded  by  the European  Commission  under  the Horizon 2020 Framework Programme. We would like to thank Daniele Santamaria for his support regarding the OASIS ontology.
\end{acknowledgements}

\bibliographystyle{vancouver}
\bibliography{r2r}

\end{document}